\documentclass[preprint,showpacs,preprintnumbers,amsmath,amssymb]{revtex4}
\usepackage{graphicx}% Include figure files
\usepackage{dcolumn}% Align table columns on decimal point
\usepackage{bm}% bold math
\begin{document}

\title{Isospin dependence of incompressibility in relativistic and non-relativistic mean field calculations}
\author{Hiroyuki Sagawa~$^{\rm a}$, Satoshi Yoshida~$^{\rm b}$, 
 Guo-Mo Zeng$^{\rm c}$, Jian-Zhong Gu$^{\rm d}$ and Xi-Zhen Zhang$^{\rm e}$ }
\affiliation{$^{\rm a}$~Center for Mathematical Sciences, 
The University of Aizu \\
Aizu-Wakamatsu, Fukushima 965-8580, Japan\\
$^{\rm b}$~Science Research Center, Hosei University \\
2-17-1 Fujimi, Chiyoda, Tokyo 102-8160, Japan \\
$^{\rm c}$ School of Physics, Jilin University, Changchun
130021, P. R. China\\
$^{\rm d}$ China Institute of Atomic Energy, P.O. Box 275(18),
Beijing 102413, P. R. China\\
$^{\rm e}$ China Institute of Atomic Energy, P.O. Box 275(18),
Beijing 102413, P. R. China}
\date{\today}

\begin{abstract}
The isospin dependence of incompressibility is investigated in the 
Skyrme Hartree-Fock (SHF) and relativistic mean field (RMF) models.
The correlations between the nuclear matter incompressibility and 
the isospin dependent term of the  finite nucleus incompressibility 
is elucidated by using 
the Thomas-Fermi approximation. 
The Coulomb term is also studied by using various different Skyrme
 Hamiltonians and RMF Lagrangians.  
The symmetry energy coefficient of incompressibility is extracted 
to be $K_{\tau}=-(500\pm50)$ MeV from the recent experimental data of 
 isoscalar giant monopole resonances (ISGMR) in Sn isotopes.   
 Microscopic HF+random phase approximation (RPA) calculations 
 are also performed with Skyrme interactions for 
 $^{208}$Pb and Sn isotopes to study the strength distributions of ISGMR.
.
\end{abstract}

\bigskip
\pacs{PACS:21.60.-n, 21.65.+f}
%\narrowtext
\maketitle
%******************
\section{Introduction}
%******************

The study of the correct value of the nuclear matter 
incompressibility $K_{\infty}$ 
continues to be an active area both theoretically and experimentally.
The value of $K_{\infty}$ is the most fundamental quantity used in determining 
the nuclear matter equation of states (EOS). 
The important experimental information on $K_{\infty}$ is
provided by the isoscalar giant monopole resonances (ISGMR)
in finite nuclei.  Non-relativistic and relativistic
mean-field models have been successfully applied
% in providing microscopic 
descriptions 
 of many properties of the nuclear ground states and also collective 
excitations including giant resonances.
The Skyrme Hartree$-$Fock (SHF) model is one commonly used non-relativistic 
mean field model \cite{Skyrme,Vautherin}.  The Hartree-Fock(HF)+random phase
 approximation (RPA) calculations with Skyrme interactions 
 were also performed to obtain 
 the response functions
of ISGMR \cite{Blaizot,HS97}. 
The relativistic mean field (RMF) model is based on an effective Lagrangian 
for the interacting many-body system \cite{Serot3}. 
 The time-dependent Hartree(TDRMF) and
RPA calculations were performed for the
ISGMR \cite{Piek02,Vret03} based on the same RMF Lagrangian.

The nuclear matter incompressibility $K_{\infty}$ is determined by the second 
 derivative of the energy per particle $E/A$ with respect to the density 
$\rho$ at the saturation point,  
%%%%% Nov. 1st 2006%%%%%%
\begin{equation}
 K_{\infty}=9\rho^2\frac{d^2}{d\rho^2} \left. \left(\frac{h}{\rho} \right) 
\right|_{\rho=\rho_{nm}},
\label{eq:Knm}
\end{equation}
where $h$ is the isoscalar part of the Hamiltonian density $H_{nm}$ for nuclear 
matter.
The nuclear matter incompressibility $K_{\infty}$ is not a directly 
measurable quantity. Instead, the energy of ISGMR $E_{ISGMR}$ is expressed 
in terms of the finite nucleus incompressibility $K_A$ as  \cite{Blaizot}
\begin{equation}
E_{ISGMR}=\sqrt{\frac{\hbar^2K_A}{m<r^2>_m}},
\label{eq:E_ISGMR}
\end{equation}
where m is the nucleon mass and $<r^2>_m$ is the mean square 
mass radius of the ground state.  The finite nucleus incompressibility 
can be parameterized by means of a similar expansion 
to the liquid drop mass formula with the volume, surface, symmetry and Coulomb 
 terms; 
\begin{equation}
K_A=K_{\infty}+K_{surf}A^{-1/3}+K_{\tau}\delta^2+K_{Coul}\frac{Z^2}{A^{4/3}},
\label{eq:K_A}
\end{equation}
where $\delta=(N-Z)/A$.

In this work, we study correlations among nuclear matter 
and finite nucleus incompressibilities for a large number of different Skyrme and RMF parameter 
sets.  In particular, we study the correlations between $K_{\infty}$ and 
$K_{\tau}$, and also between  $K_{\infty}$ and $K_{Coul}$.  Then, we 
extract the values of $K_{\infty}$ and $K_{\tau}$ from experimental ISGMR 
recently observed in RCNP, Osaka University, and also in Texas A\&M University.
Skyrme HF+RPA calculations are also performed to study the detailed structure of 
ISGMR taking into account the coupling to the continuum. In section II, we
 describe the nuclear matter and finite nucleus incompressibilities by using 
the Thomas-Fermi approximation.   HF+RPA results of  ISGMR  will be shown 
for  $^{208}$Pb and Sn isotopes  in Section III.  
The summary is given in section IV.

\section{Thomas-Fermi approximation for Skyrme and RMF Hamiltonian density}

The Skyrme interactions used in many recent studies have nine 
parameters ($t_0,t_1,t_2,t_3,x_0,x_1,x_2,x_3,\alpha)$ 
in addition to the parameters of the spin-orbit interaction.
The energy density functionals of the Skyrme interaction 
are expressed  by using the Thomas-Fermi approximation
for the kinetic energy density.
The nuclear matter properties are defined by using these energy density 
functionals. Various correlations among nuclear matter properties have been
discussed in the cases of the SHF and RMF models recently 
\cite{Yoshida06,yoshida,Brown,Brown2}.
It was shown that the Skyrme parameters can be 
 expressed analytically in terms of 
the isoscalar and the isovector nuclear matter properties of the Hamiltonian 
density in ref. \cite{Yoshida06}.

In this section, we detail study of correlations between
 the nuclear 
matter incompressibility $K_{\infty}$ and the symmetry term $K_{\tau}$ of the
finite nucleus incompressibility, and also between $K_{\infty}$ and the Coulomb
 term $K_{Coul}$ 
 in the SHF model for 14 different parameter sets (SI, SIII, SIV, SVI, 
Skya, SkM, SkM$^{*}$, SLy4, MSkA, SkI3, SkI4, SkX, SGI, SGII) taken from Refs. 
\cite{Vautherin,Beiner,Kohler,Bohigas,Bartel,Reinhard,Chabanat,Sharma,Brown,Giai} 
and the RMF model for seven different parameter sets (NL3, NLSH, NLC, TM1, TM2, 
DD-ME1, DD-ME2) taken from refs. \cite{Lalazissis,Sharma2,Serot3,Toki,Nik,Lala}. The non-linear potential of $\sigma$ mesons are introduced in the first five
parameter sets of RMF.  The non-linear potential of the $\omega$ meson is also added 
in the parameter sets TM1 and TM2.  
The meson$-$nucleon couplings depend on the total density in DD-ME1 and 
DD-ME2 parameter sets, while the couplings are constants in the other five 
parameter sets.
The non-linear potential of $\sigma$ 
mesons are fixed to be zero in DD-ME1 
and DD-ME2 parameter sets.

The isoscalar part $h(\rho)$ and the isovector part 
 $\varepsilon_{\delta}(\rho)$ 
of the Hamiltonian density $H_{nm}$ are defined by 
\begin{eqnarray}
h(\rho)&=&\lim_{I \rightarrow 0} H_{nm}, \\ 
\varepsilon_{\delta}(\rho)&=&\frac{1}{2}~\lim_{I \rightarrow 0}~
\frac{\partial^2}{\partial I^2}~\left( \frac{H_{nm}}{\rho} \right), 
\end{eqnarray}
where $I$ is the asymmetry parameter $I=(\rho_{n}-\rho_{p})/\rho$.
The derivative terms and the Coulomb term in the Hamiltonian density 
do not give any contribution in the 
infinite nuclear matter calculations.
The Thomas-Fermi approximation can be applied  for the kinetic energy of the 
Hamiltonian density $H_{nm}(\rho_{n},\rho_{p})$ of nuclear matter in the SHF 
and RMF model.  The explicit forms of the Hamiltonian densities $H_{nm}$ are 
found in ref. \cite{Yoshida06}.

The physical properties of infinite symmetric nuclear matter with 
$\rho_{n}=\rho_{p}$ can be obtained from the following six equations: 
\begin{eqnarray}
0&=&\left. \frac{\partial}{\partial \rho} \left( \frac{h}{\rho} \right) 
\right|_{\rho=\rho_{nm}}, \label{rhonm} \\ 
-E_{0}&=&\frac{h(\rho_{nm})}{\rho_{nm}}~~~~~~~~~~~~(~{\rm in~SHF}~), \nonumber \\ 
-E_{0}&=&\frac{h(\rho_{nm})}{\rho_{nm}}-M~~~~~(~{\rm in~RMF}~), \label{E0} \\ 
K_{\infty}&=&\left. 9 \rho^{2} \frac{\partial^2}{\partial \rho^2} \left(\frac{h}{\rho} 
\right) \right|_{\rho=\rho_{nm}}, \label{K} \\ 
J&=&\varepsilon_{\delta}(\rho_{nm}), \label{J} \\ 
L&=&\left. 3 \rho \frac{\partial}{\partial \rho} \varepsilon_{\delta}(\rho) 
\right|_{\rho=\rho_{nm}}, \label{L} \\ 
K_{sym}&=&\left. 9 \rho^2 \frac{\partial^2}{\partial \rho^2} 
\varepsilon_{\delta}(\rho) \right|_{\rho=\rho_{nm}}, \label{Ksym}
\end{eqnarray}
where $\rho_{nm}$, $E_{0}$, $K_{\infty}$ and $J$ are
 the nuclear saturation density, the 
binding energy per nucleon, the incompressibility of symmetric nuclear matter 
and the symmetry energy, respectively.

We denote $K_{\tau}$ the symmetry term of the finite nucleus 
 incompressibility 
$K_{A}$  because the symbol $K_{sym}$ has been already used 
as one of the isovector nuclear matter properties defined by Eq.(\ref{Ksym}).
 The volume 
term of the finite nucleus incompressibility $K_{A}$  is identified as 
the nuclear matter incompressibility $K_{\infty}$.   
The symmetry contribution $K_{\tau}$ is related to nuclear matter properties 
as \cite{Blaizot}, 
\begin{equation}
\left. K_{\tau}=K_{sym}+3 L -\frac{27 L \rho_{nm}^{2}}{K_{\infty}} 
\frac{d^{3} {h}}{d \rho^{3}} \right|_{\rho=\rho_{nm}}.
\label{Ka-sym-eq}
\end{equation}
The analytic formulas for $K_{surf}$ and $K_{Coul}$ are given by
\begin{eqnarray}
K_{surf}=4\pi\ {r_0}^2 \left[ 4 \sigma(\rho_{nm}) +9 \rho_{nm} \left. 
\frac{d^2 \sigma}{d \rho^2} \right|_{\rho=\rho_{nm}} 
+\frac{54 \sigma(\rho_{nm}) \rho_{nm}^2}{K_{\infty}} \left. 
\frac{d^3 h}{d \rho^3} \right|_{\rho=\rho_{nm}} \right],
\label{Ksurf}
\end{eqnarray}
\begin{eqnarray}
\left. K_{Coul}=\frac{3}{5}\frac{e^2}{r_0}\left(1-\frac{27\rho_{nm}^2}{K_{\infty}}
\frac{d^{3} {h}}{d \rho^{3}} \right|_{\rho=\rho_{nm}}\right),  
\end{eqnarray}
where $r_{0}$ is the radius constant defined by
\begin{equation}
r_{0}= \left(\frac{3}{4\pi\rho_{nm}}\right)^{1/3}.
\end{equation}
In Eq.(\ref{Ksurf}), $\sigma$ is a surface tension in symmetric semi-infinite nuclear matter 
defined by
\begin{equation}
\sigma(\rho_{nm})=\int_{-\infty}^{\infty} \left[ H(\rho)-
\frac{h(\rho_{nm})}{\rho_{nm}} \rho \right] 
dz,
\end{equation}
where $H$ is the Hamiltonian density.
$K_{surf}$ can be evaluated by the extended Thomas-Fermi approximation and 
the scaled HF calculations on semi-infinite nuclear matter in the SHF model.
These evaluations show an approximate relation $K_{surf}\sim -K_{\infty}$   
within an accuracy of a few \% in the SHF model.  In RMF, the study of 
an extended Thomas-Fermi approximation gives a slightly larger surface contribution, 
for example,  $K_{surf}\sim -1.16K_{\infty}$ in the case of NL3.

\begin{figure}[h]
\includegraphics[width=5.5in,clip]{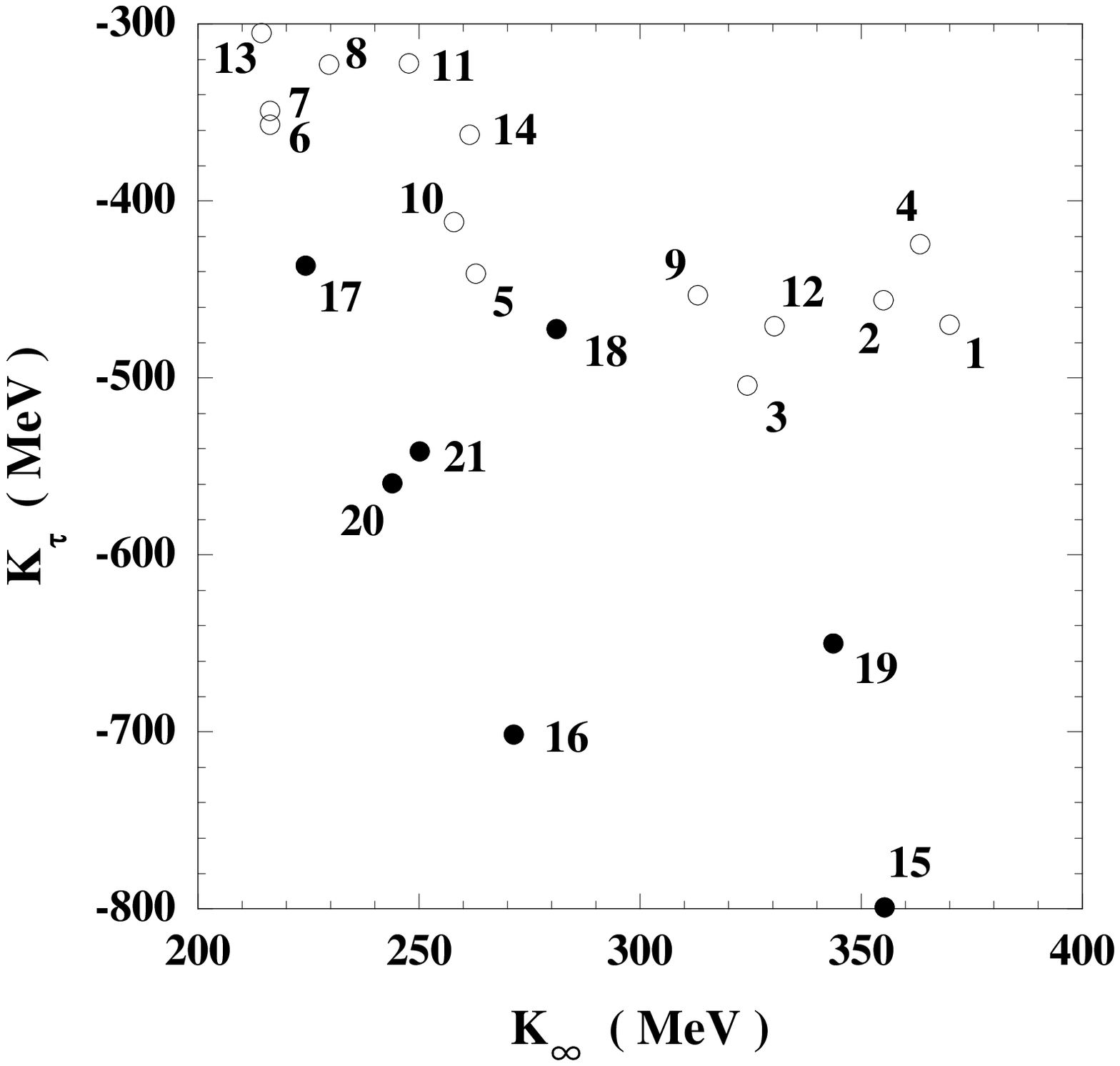}
\caption{Correlation between the nuclear matter  incompressibility $K_{\infty}$  and  the symmetry  contribution  $K_{\tau}$ in the finite nucleus 
incompressibility calculated by using 14
parameter sets of the SHF (open circles) and seven parameter sets of RMF (filled 
circles).
The numbers denote the different parameter sets: 1 for SI, 2 for SIII, 3 for 
SIV, 4 for SVI, 5 for Skya, 6 for SkM, 7 for SkM$^{*}$, 8 for SLy4, 9 for MSkA, 
10 for SkI3, 11 for SkI4, 12 for SkX, 13 for SGII, 14 for SGI, 15 for 
 NLSH, 16 for NL3, 17 
for NLC, 18  for TM1, 19 for TM2, 20 for DD-ME1 and 21 for DD-ME2.
}
\label{fig:Ktau}
\end{figure}

\begin{figure}[p]
\includegraphics[width=5.5in,clip]{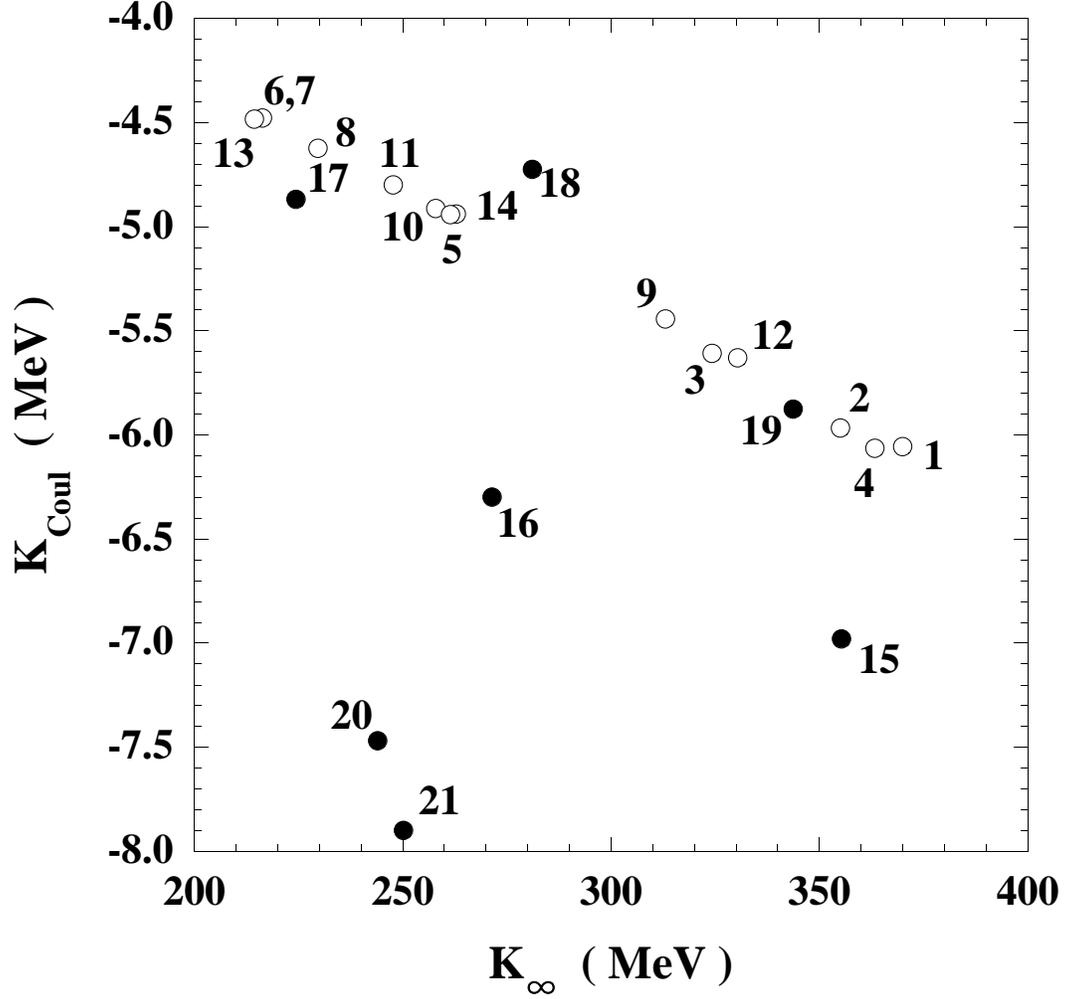}
\caption{Correlation between the nuclear matter  incompressibility $K_{\infty}$  and  the Coulomb   contribution  $K_{Coul}$ in the finite nucleus 
incompressibility  calculated by using 14 
parameter sets of the SHF (open circles) and seven parameter sets of RMF (filled 
circles).
See the caption of Fig.\ref{fig:Ktau} and the text for details.
}
\label{fig:Kcoul}
\end{figure}

The values of $K_{\tau}$ and  $K_{Coul}$ are calculated by using
various Skyrme Hamiltonians and RMF Lagrangians and 
shown in Figs. \ref{fig:Ktau} and  \ref{fig:Kcoul}.
$K_{\tau}$  is largely negative and has anti-correlation with the 
nuclear matter incompressibility $K_{\infty}$.  Namely, any Hamiltonian 
which has a larger $K_{\infty}$  gives a smaller $K_{\tau}$.
The variations  of   $K_{\tau}$ for the Skyrme interactions are
\begin{equation}
K_{\tau} =(-400\pm100) {\rm ~MeV}   \hspace{2cm} \mbox{for Skyrme interaction}.
\end{equation}
On the other hand, the values of RMF are largely negative and 
have  more variation among the seven effective Lagrangians,
\begin{equation}
K_{\tau} =(-620\pm180) {\rm ~MeV}   \hspace{2cm} \mbox{for RMF Lagrangian}.
\end{equation}
 In principle,
the value $K_{Coul}$ should be model-independent.  Nevertheless,
we can see a  weak correlation between $K_{\infty}$ and  $K_{Coul}$ 
in Fig.  \ref{fig:Kcoul}.  The correlation between 
 $K_{\infty}$ and  $K_{Coul}$ can be expressed analytically by using the 
isoscalar nuclear matter properties as given in the Appendix. Among the 13 
parameter sets of Skyrme interactions, the variation of $K_{Coul}$ 
 is rather small,
\begin{equation}
  K_{Coul} =(-5.2\pm0.7) {\rm ~MeV}   \nonumber
\end{equation}
compared with that of $K_{\tau}$.  The values of $K_{Coul}$ in RMF show
essentially the same trend, but have a larger variation.

 Recently, the ISGMR strength distributions in the Sn isotopes from $^{112}$Sn 
to  $^{124}$Sn have been
measured by using inelastic $\alpha$ scatterings at RCNP, Osaka 
University \cite{Umesh}.  
 The ISGMR of  Sn isotopes were also studied in Texas A \& M 
 University \cite{Youngblood,Lui2}.  
 The  value of $K_{\tau}$ was extracted to be
%\begin{equation}
$K_{\tau} =(-395\pm40)$~MeV  for Sn isotopes  
% \hspace{2cm} \mbox{for Sn isotopes},
%\end{equation}
assuming  $K_{A}$ to be a quadratic relation with the symmetry 
parameter $\delta=\frac{N-Z}{A}$, i.e.,  $K_{A}$=c+$K_{\tau}\delta^2$ where
$c$ is a constant.  We should be careful to apply 
this relation to obtain the value $K_{\tau}$ 
%This relation is somewhat questionable 
since the surface
 and the Coulomb contributions in Eq. (\ref{eq:K_A}) are  also 
 functions of the mass number $A$.  
We examine the $K_{\tau}$ dependence of the $E_{ISGMR}$ by using the formula
 (\ref{eq:K_A}).
In Fig. 3, the difference of the compressibility $\Delta K_A=K_A-K_{A=112}$ 
is plotted as a function of $\delta=(N-Z)/A$.  We adopt 
 four Skyrme interactions and two 
RMF Lagrangians.   In the analysis, the surface term is taken to be 
  $K_{surf}= -K_{\infty}$ for 
 Skyrme model and  
 $K_{surf}= -1.18K_{\infty}$ for RMF model. The adopted interactions vary from a smaller 
$K_{\tau}$ value of -350 MeV for SkM$^*$ to a larger value of
-700 MeV for NL3. The empirical isospin dependence of $\Delta K_A$ is close
%well
%reproduced by
 to the results of  SIII, SIV and DD-ME1, which have $K_{\tau}=-(500\pm50)$
 MeV. The present 
extracted value is 
about 30\%  larger than the value reported in ref. \cite{Umesh}.
  This difference is mainly due to 
the mass number dependence of $K_{surf}$ and 
$K_{Coul}$ which were neglected in ref.  \cite{Umesh}.  

\section{HF+RPA calculations for ISGMR}

In order to extract the incompressibility $K_{\infty}$ for infinite 
nuclear matter from the experimental values of the ISGMR, 
self-consistent RPA 
 calculations were performed and results thereof were compared directly with the experimental 
 data both 
in SHF and RMF model \cite{Blaizot,HS97,Piek02,Vret03}.  
 The sum rule approach with the constrained HF calculations
 was also adopted to disentangle 
$K_{\infty}$ in ref. \cite{Colo}. 
%The correlation between the symmetry 
% contribution $K_{\tau}$ of the finite  nucleus
%incompressibility $K_{A}$  and the symmetry energy $J$ was also 
%discussed in \cite{Colo}.
The experimental data of $^{208}$Pb was adopted in these studies since 
ISGMR data is most well-established in this nucleus. 
%  the energy of ISGMR $E_{ISGMR}=\sqrt{\frac{m_1}{m_{-1}}}$ in 
%$^{208}$Pb where 
%$m_k$ is k-th energy weighted sum rule value (EWSR).  
There has been an attempt to extract  
the volume, surface, symmetry, and Coulomb 
terms in Eq. (\ref{eq:K_A}) 
 by using various sets of
 experimental data of ISGMR in different nuclei \cite{SY}.
However, the results depended very much on the adopted data set and gave 
a negative conclusion to the feasibility of simultaneously determining the four terms of Eq. (\ref{eq:K_A}) from experimental data.  
In this study, we perform HF+RPA calculations for $^{208}$Pb, 
$^{112}$Sn,  $^{116}$Sn, $^{120}$Sn and $^{124}$Sn to enable quantitative 
comparisons to be made with the experimental strength distributions of ISGMR and
 try to extract realistic values of  $K_{\infty}$ and  $K_{\tau}$.

 The RPA strength 
distribution  
\begin{equation}
 S(E)=\sum_n|<n|Q|0>|^2\delta(E-E_n)
\label{eq:strength}
\end{equation}
is calculated by using the IS monopole operator
\begin{equation}
 Q^{\lambda=0,\tau=0}=\frac{1}{\sqrt{4\pi}}\sum_ir^2_i .
\label{operator}
\end{equation}
%The excitation energy of ISGMR will be obtained  by 
%using 
The k-th energy moment of the transition strength is defined by
\begin{equation}
m_k=\int dE E^k S(E).
\label{eq:moment}
\end{equation}
The average energy will be obtained by
a ratio between the moments $m_1$ and $m_0$, 
\begin{equation}
 \bar{E}=m_1/m_0.
\label{eq:aveex}
\end{equation}
The scaling model of ISGMR gives the excitation energy
\begin{equation}
 E_s=\sqrt{m_3/m_1},
\label{eq:exs}
\end{equation}
while the excitation energy produced by the constrained HF model is written as
\begin{equation}
 E_c=\sqrt{m_1/m_{-1}}.
\label{eq:exc}
\end{equation}
The excitation energies defined by Eqs. (\ref{eq:aveex})-(\ref{eq:exc}) are
identical in the case of  a sharp single peak 
which exhausts 100\% of the sum rule. 
  However, in reality, both the experimental 
data and the calculated results show a large width of a few MeV even in the 
most well-established ISGMR in  $^{208}$Pb.  In particular, the scaling 
energy $E_s$ has a
large uncertainty due to the high energy tail of monopole strength, which is 
always the case in experimental data.  On the other hand, $\bar{E}$ and 
 $E_c$ are rather close within a 0.1$\sim$0.2 MeV difference even when the 
ISGMR peak has a large width.  Because of these reasons and also due to the 
theoretical clear background, we identify  $E_c$ as  $E_{ISGMR}$ to obtain
$K_A$ through Eq. (2).  It should be noticed that the  rms mass radius is 
needed to obtain $K_A$ through Eq. (2), but not the charge radius.  Since the
mass radii of Sn isotopes were not   
determined experimentally so far, we use HF mass radii to analyse 
  $K_A$.

%A measure of the width is given by the variance
%\begin{equation}
%\sigma=\sqrt{m_2/m_0-(m_1/m_0)^2}
%\label{eq:variance}
%\end{equation}

 The continuum HF+RPA results for ISGMR in $^{208}$Pb 
are shown in Fig. 4 with three different
 Skyrme interactions SIII, SGI and SkM$^*$.  The incompressibility of 
 The SIII interaction is 355 MeV, while those of SGI and SkM$^*$ are 256 MeV and
 217 MeV, respectively. Various excitation energies (\ref{eq:aveex}), 
 (\ref{eq:exs}) and  (\ref{eq:exc}) are given in Table \ref{tab:pb-ismn}. 
  As far as the excitation energy is concerned, 
SkM$^*$ shows a fairly good agreement 
 with the experimental data obtained by
 ($\alpha,\alpha ')$ scatterings \cite{Youngblood}.  
The average energies $\bar{E}$
  of SGI and SIII are
 1.0 MeV and 3.3 MeV higher than the empirical one, respectively. 
  The calculated width for the SkM$^*$ interaction 
 shows almost the same width as that of experimental data.
  This agreement implies that the dominant contribution of 
 the width of ISGMR stems 
  from the Landau damping and the coupling to the continuum, which are properly 
taken into account in the present calculations.  On the 
other hand, the coupling to the 
 many-particle many-hole states might have a minor effect on the width of 
ISGMR.  The 
 agreement between the results of SkM$^*$ and the experimental data 
  strongly suggests a lower incompressibility 
 $K_{\infty}\simeq$220 MeV as a realistic one for infinite nuclear matter. 
% which is consistent with the constrained HF
% study in ref.  \cite{Colo}.  
The two-body spin-orbit and the two-body Coulomb interactions 
 are not taken into account 
in the present RPA calculations although the HF calculations include both 
interactions.  
 It was pointed out in ref. \cite{Shlomo} that the net effect of the
 two interactions in RPA decrease the centroid energy of ISGMR E$_{ISGMR}$ 
  in $^{208}$Pb by
about 300 keV.   Then, 
%  decrease   using 
an approximate relation $(\delta K_{\infty}/K_{\infty})$
=2($\delta $E$_{ISGMR}$/E$_{ISGMR}$) from Eq. (2) gives a decrease
 of the incompressibility by $\delta K_{\infty}\sim$10 MeV. Due to 
this effect, a realistic incompressibility should be slightly larger than
than of SkM$^*$ to be  $K_{\infty}\simeq$230 MeV.
In ref. \cite{Colo}, 
 the nuclear matter
 incompressibility $K_{\infty}$ was discussed 
 taking into account the full interactions of 
 various Skyrme parameter sets in the constrained HF model.  
Comparing the calculated excitation energies 
 $E_c$ in Eq. (\ref{eq:exc})  with
 the experimental data of
 ISGMR in $^{208}$Pb, they concluded that 
  the nuclear matter
 incompressibility  should be $K_{\infty}$=230-240 MeV in order to explain the
 experimental data.   This conclusion  is consistent with 
the present HF+RPA calculations with SkM$^*$ 
 since the two-body spin-orbit and 
Coulomb interactions give the net effect of
  $\delta K_{\infty}\sim$10 MeV. 

\begin{table}[htb]
\caption{\label{tab:pb-ismn}
  The energies of ISGMR in $^{208}$Pb and the finite nucleus incompressibility
  $K_A$ calculated by using Eq. (\ref{eq:K_A}) with $E_{ISGMR}=E_c$. 
   The HF+RPA results are calculated by using SkM$^*$, SGI and SIII
   interactions.
 The sum rules $m_k$ are obtained by summing up the strength
up to E$_x$=25.5 MeV. 
The average energy is defined by the ratio $\bar{E}=m_1/m_0$ in Eq. 
(\ref{eq:aveex}), while $E_s$ and $E_c$ are obtained as 
 $E_s=\sqrt{m_3/m_1}$ and
$E_c=\sqrt{m_1/m_{-1}}$ in Eqs. (\ref{eq:exs}) and (\ref{eq:exc}). 
%while the variance $\sigma$ is calculated by using Eq. (\ref{eq:variance}).
 The experimental data are taken from ref. 
 \cite{Youngblood}.  The experimental value $K_A$ is calculated by using
 $E_c$ and the mass radius of the SkM$^*$ interaction.
 All values are given in units of  MeV.}
\vspace{0.3cm}
\begin{center}
\begin{tabular}{c|c|c|c|c|c} \hline
   & $\bar{E}$  &  $E_s$ &  $E_c$ & $K_A$ & $K_{\infty}$\\\hline
 SkM$^*$ &  13.87 & 14.29 & 13.79  &141.5 & 217  \\\hline
 SGI &   15.17 & 15.47 & 15.07   &170.0 & 256 \\\hline
 SIII &  17.47 & 17.88 & 17.31  &226.4 & 355 \\\hline
 exp (Texas A\&M) &  14.17$\pm$0.28 & ----- & 14.18$\pm$0.11 &  149.6$\pm$2.3  &     \\\hline
  \end{tabular}
\end{center}
\end{table}

Next, let us discuss the isospin dependence of the excitation energies of 
ISGMR in Sn isotopes.  
For the HF+RPA calculations of Sn isotopes, we adopt the SkM$^*$ 
interaction,  discretizing the continuum with 
a large harmonic oscillator basis up to the maximum 
major quantum number $N_{max}$=16.  
The RPA calculations are performed by using the filling 
approximation for the neutron orbits, i.e., the neutrons occupy
 the orbits from the bottom of the 
potential to the Fermi level in order. The last neutron orbit is
partially filled according to the neutron number.
The calculated values of the transition strength  $ B(E)$
 are averaged by using a weighting factor $\rho(E_x-E)$
\begin{equation}
  S(E_x)= \sum_i B(E_i) \rho(E_x-E_i),
\end{equation}
where 
\begin{equation}
\rho(E_x-E_i)=\frac{1}{\pi}\frac{\Delta/2}{(E_x-E_i)^2+(\Delta/2)^2}.
\end{equation}
The width parameter is taken to be $\Delta$=1 MeV in Fig. \ref{fig:sn-mono}.
We also performed the continuum RPA calculation in a nucleus $^{116}$Sn 
 and found essentially identical results as far as the excitation energies 
listed in Table \ref{tab:energy} are concerned. 
As seen in Fig. \ref{fig:sn-mono}, the RPA results show reasonable 
agreement with the experimental data obtained by Texas A\&M \cite{Youngblood,Lui2} and also by RCNP \cite{Umesh}.  The various average energies  $\bar{E}=m_1/m_0$,  $E_s=\sqrt{m_3/m_1}$ and
$E_c=\sqrt{m_1/m_{-1}}$ are 
listed in Table \ref{tab:energy}. In general, the calculated average 
energies by RPA with SkM$^*$ are few hundreds keV higher than
 the empirical ones.  The two sets of recent experimental data 
 show slight differences 
 in $^{112}$Sn and $^{124}$Sn which should be confirmed in the future by further 
experimental study.  The calculated excitation energies decrease from 
 $^{112}$Sn and $^{124}$Sn by about 1 MeV which is consistent with the 
observed data.  This decrease is expected from a large negative symmetry term 
$K_{\tau}$ in the finite nucleus incompressibility discussed in Section 2.

The RPA results give about 80\% of the observed widths of ISGMR in Sn 
isotopes.  These  results are almost identical to the case of $^{208}$Pb.
Thus, we can conclude from the results of Sn isotopes and $^{208}$Pb that 
the major part of the width of ISGMR stems from the Landau damping and 
the coupling to the continuum.

\begin{table}[htb]
\caption{\label{tab:energy}
  The energies of ISGMR in Sn isotopes.  
   The HF+RPA results are obtained by using the SkM$^*$ interaction.
 The RPA sum rules $m_k$ are obtained by summing up the strength
up to E$_x$=25.5 MeV to be consistent with experimental 
data. 
The average energy obtained by the ratio $\bar{E}=m_1/m_0$  in Eq. 
(\ref{eq:aveex}), while $E_s$ and $E_c$ are defined by $E_s=\sqrt{m_3/m_1}$ and
$E_c=\sqrt{m_1/m_{-1}}$. 
%while the variance $\sigma$ is calculated by using Eq. (\ref{eq:variance}).
 The experimental data are taken from refs. 
 \cite{Youngblood,Lui2}(Texas A\&M) and \cite{Umesh}(RCNP). 
 The experimental value $K_A$ is calculated by using
 $E_c$ and the mass radius of the SkM$^*$ interaction. 
  All  values are given in units of  MeV.}
\vspace{0.3cm}
\begin{center}
\begin{tabular}{c|c|c|c|c|c|c|c|c|c|c|c} \hline
 &\multicolumn{4}{c|} {RPA(SkM$^*$)} & \multicolumn{3}{c|}{Texas A\&M} & \multicolumn{4}{c}{RCNP} \\\hline
 &$\bar{E}$  & $E_s$ & $E_c$& $K_A$  &$\bar{E}$  & $E_s$ & $E_c$   &$\bar{E}$  & $E_s$ & $E_c$  & $K_A$  \\\hline 
%          & (MeV) &   (MeV) & (MeV)  & &  (MeV) (keV) &  (fm$^2$) \\\hline \hline
 $^{112}$Sn & 17.1 & 17.5 & 16.9 & 142.7& 15.43 & 16.05   & 15.23 & 16.2 & 16.7 & 16.1 & 129.5 $\pm$1.6   \\\hline
 $^{116}$Sn & 16.6 & 17.0  & 16.5 & 140.2 &16.07  & ---  &15.90 & 15.8 & 16.3 &  15.7  & 126.9  $\pm$1.6   \\\hline
 $^{120}$Sn & 16.1 & 16.5 & 15.9 &133.7 &---   &---  &--- & 15.7  & 16.2 & 15.5  &127.0  $\pm$1.7  \\\hline
 $^{124}$Sn & 16.2 & 16.7 & 16.1 & 137.1& 14.50 & 14.96 & 14.33 & 15.3 & 15.8 & 15.1 & 122.7  $\pm$1.7\\\hline
  \end{tabular}
\end{center}
\end{table}

%******************
\section{Summary}

In summary, we studied the ISGMR of $^{208}$Pb and Sn isotopes in order to 
disentangle the nuclear matter incompressibility $K_{\infty}$ and also 
the symmetry term $K_{\tau}$ in the finite nucleus incompressibility.
Firstly, the Thomas Fermi approximation is adopted to obtain the various 
terms of the finite nucleus incompressibility $K_A$ in the SHF and RMF models.
The correlations between  $K_{\infty}$  and  $K_{\tau}$ and also between
 $K_{\infty}$ and  $K_{Coul}$ are elucidated in the various sets of the 
Skyrme Hamiltonians and RMF Lagrangians.  We extracted the symmetry 
term to be  $K_{\tau}=-(500\pm50)$ MeV from the analysis of the isospin 
dependence of the  
excitation energies of ISGMR in Sn isotopes.  Secondly, we perform HF+RPA 
calculations to study detailed structure of ISGMR in $^{208}$Pb and Sn isotopes.  In $^{208}$Pb, the results of continuum RPA with
three sets of Skyrme interactions SkM$^*$, SGI and SIII are compared with
 the experimental data.  It is shown that SkM$^*$ gives a 
satisfactory description of ISGMR of $^{208}$Pb in both the excitation 
energy and the width. The nuclear matter incompressibility is 
extracted to be $K_{\infty}\sim$230 MeV from the RPA analysis with the 
additional effects of the two-body spin-orbit and Coulomb interactions.
The isospin dependence of ISGMR in Sn isotopes is also studied by using 
the HF+RPA calculations with the SkM$^*$ interaction.  We pointed out that 
the RPA results give a reasonable account of the isospin dependence of the
 observed data, although the calculated excitation energies are few 
hundreds of keV higher than the observed ones.  It should be noted that  
 appreciable differences exist between the two measurements in the excitation energies of $^{112}$Sn and $^{124}$Sn,  listed in Table
 \ref{tab:energy}, 
 which makes it difficult to accurately determine
the value 
 $K_{\tau}$ of  the finite nucleus incompressibility. 
This problem remains 
 as a future challenge both in experimental and theoretical 
studies.

\begin{acknowledgments}
We thank  Y.-W. Lui, U. Garg and M. Fujiwara for fruitful discussions and 
also for providing their experimental data. 
This work is supported in part by a Grant-in-Aid for Scientific Research under 
program number (C(2)) 16540259 from the Japanese Ministry of Education, Culture, 
Sports, Science and Technology.
\end{acknowledgments}

\appendix*

\section{Analytic formulas for $K_{Coul}$}
%\begin{verbatim}

%\begin{table}[h]
\begin{table}[htb]
\caption{\label{tab:appendix}
The values of $a$ and $b$ in Eq. (A.4) as a function of the effective mass
$m^*$.}
\begin{tabular}{c|c|c}
 \hline
$m^{*}/m$&$a$&$b$ \\ \hline
0.6&$-0.0104314$&$-2.15499$ \\
0.7&$-0.0105882$&$-2.10745$ \\
0.8&$-0.0106691$&$-2.08292$ \\
0.9&$-0.0107185$&$-2.06796$ \\ \hline
\end{tabular}
\end{table}

%\section{Analytic formulas for $K_{Coul}$}
%\begin{subequations}
The Coulomb term $K_{Coul}$ can be expressed by nuclear matter 
values $K_{\infty}$ and $\rho_{nm}$, and the Hamiltonian density for symmetric 
nuclear matter $h$ as follows \cite{Yoshida06}
\begin{eqnarray}
\label{eq:a1}
K_{Coul}&=&\frac{3}{5} \frac{e^2}{r_{0}} \left( \left. 1-\frac{27 
\rho_{nm}^{2}}{K_{\infty}} \frac{d^3 h}{d \rho^3} \right|_{\rho=\rho_{nm}} 
\right) \nonumber \\
%\left. 1-\frac{27 \rho_{nm}^{2}}{K} \frac{d^3 H_{nm}}{d \rho^3} 
%\right|_{\rho=\rho_{nm}}
&=&\frac{3}{5} \frac{e^2}{r_{0}} \left(1-\frac{27}{K_{\infty}} \left[ -\frac{\hbar^2}{30 m} c \rho_{nm}^{2/3} (3 \alpha+1) 
-\frac{5}{3} (\alpha+1) E_{0} +\frac{3 \alpha+11}{27} K_{\infty} \right]\right), 
\label{eq:a2}
\end{eqnarray}
%\end{subequations}
where 
%\begin{equation}
$c=\left( 3 \pi^2/2\right)^{2/3}$ and $\alpha$ is the power of the nuclear 
density-dependent term in the Skyrme interaction \cite{Skyrme}. 
The nuclear mater incompressibility $K_{\infty}$ is also expressed as 
\begin{equation}
K_{\infty}=\frac{3 \hbar^2}{10 m} c \rho_{nm}^{2/3} \left[ 3 (3 \alpha-1) -2 (3 
\alpha -2) \left( \frac{m^{*}}{m} \right)^{-1} \right] + 9 (\alpha+1) E_{0}.
\label{eq:a3} 
\end{equation}
The linear power of the density-dependent term of the Skyrme interaction
$\alpha$ can be eliminated from Eq.(\ref{eq:a2}) by using Eq.(\ref{eq:a3});
\begin{eqnarray}
K_{Coul}&=&\frac{3}{5} \frac{e^2}{r_{0}} 
%\left[\frac{h}{10} \left( 3-2 m_{eff}^{-1} \right) +E_{0} \right]^{-1} 
\left[-\frac{K_{\infty}}{3}+3 \eta \left(-1+\frac{4}{5} m_{eff}^{-1} \right)-
2 E_{0} 
%\right.  \nonumber \\
% & &\left. 
 + \frac{3 \eta}{5 K_{\infty}} \left\{\frac{9 \eta}{10} \left( 1-m_{eff} \right)+
 E_{0} \left(27-25 m_{eff}^{-1} \right) \right\} \right] 
 \nonumber \\
& & \times \left[ \frac{\eta}{10} \left( 3-2 m_{eff}^{-1} \right) +E_{0} \right]^{-1},
\label{eq:a4}
\end{eqnarray}
where  $m_{eff}=m^{*}/m$ and $\eta=\hbar^2 c \rho_{nm}^{2/3}/m$.
%\end{equation}
%\end{eqnarray}
%\end{subequations}
%\end{verbatim}
Eq. (\ref{eq:a4}) is proportional to the power of 1,0 and -1 of $K_{\infty}$.
However, this equation can be well parameterized to be 
\begin{equation}
 K_{Coul}=aK_{\infty}+b,
\end{equation}
where $a$ and $b$ depend on the effective mass $m^*$.
With the standard value of the saturation density and the energy of 
nuclear matter $\rho_{nm}$=0.16 fm$^{-3}$ and $E_0$=16 MeV, 
the values of $a$ and $b$ are tabulated in Table A1.

\newpage

\begin{figure}[p]
\includegraphics[width=5.5in,clip]{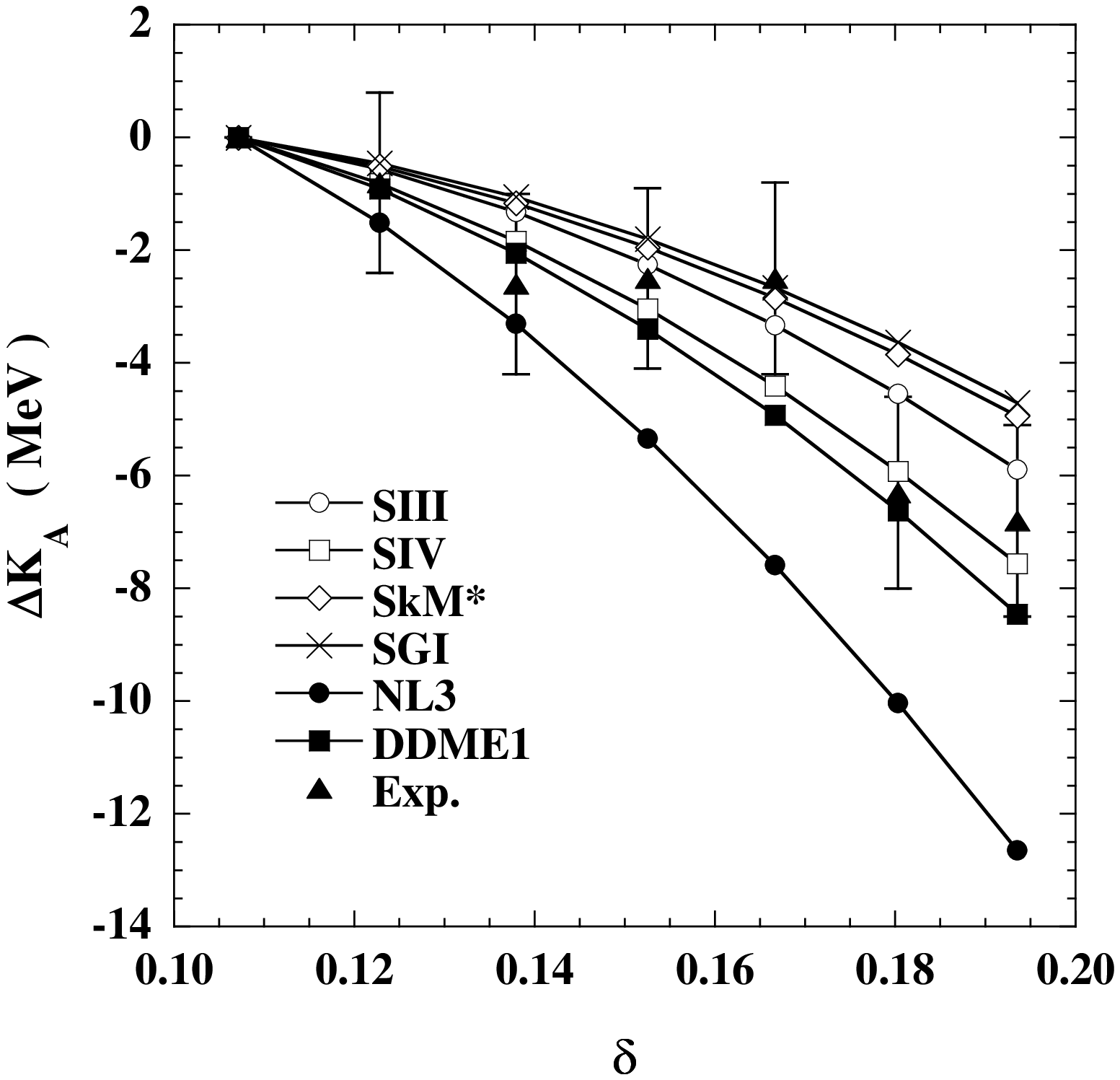}
\caption{The difference of incompressibility $\Delta K =
K_{A}-K_{A=112}$ as a function of $\delta =\frac{N-Z}{A}$. 
Experimental data are determined by using the excitation energies of ISGMR 
in   ref. \cite{Umesh} and the HF mass radii. See the text and the caption 
to Table II for details.}
\label{fig:Ka}
\end{figure}

\begin{figure}[p]
\includegraphics[width=5.5in,clip]{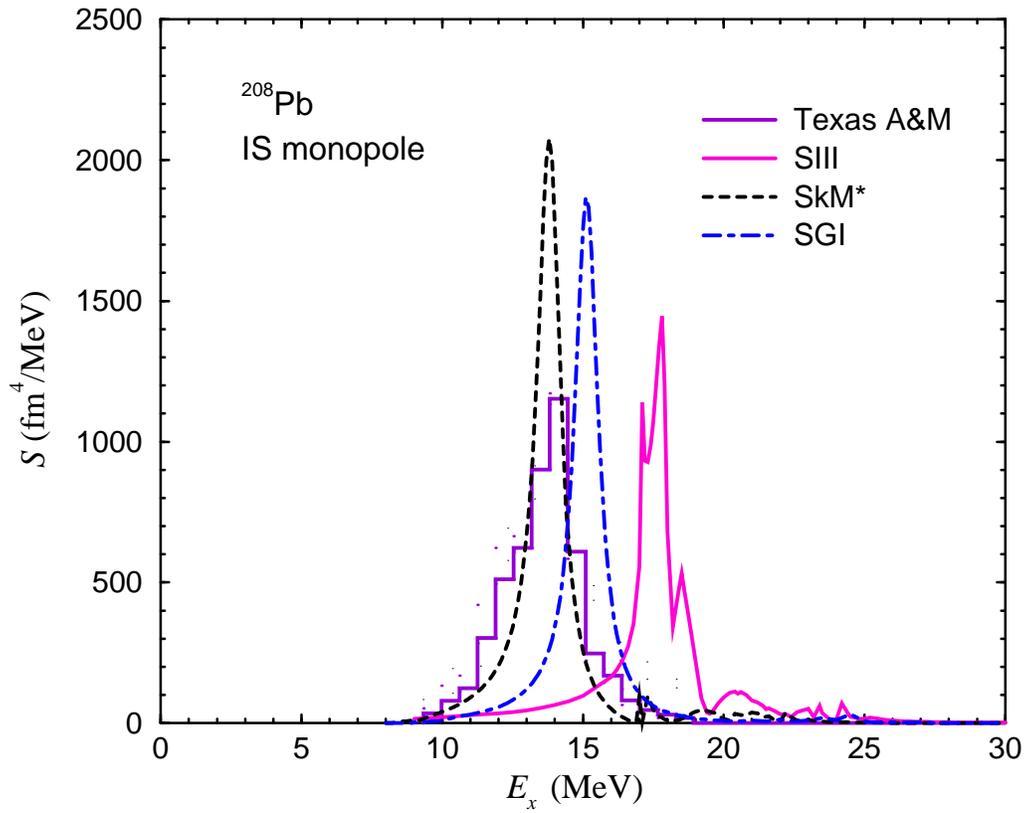}
\caption{(Color online) Continuum HF+RPA response functions of Skyrme interactions with 
the different incompressibilities SIII, SGI and SkM$^*$.
The experimental data are taken from ref. \cite{Lui}.}
\label{fig:Pb-mono}
\end{figure}

\begin{figure}[p]
\vspace{-4cm}
\includegraphics[width=3.2in,clip]{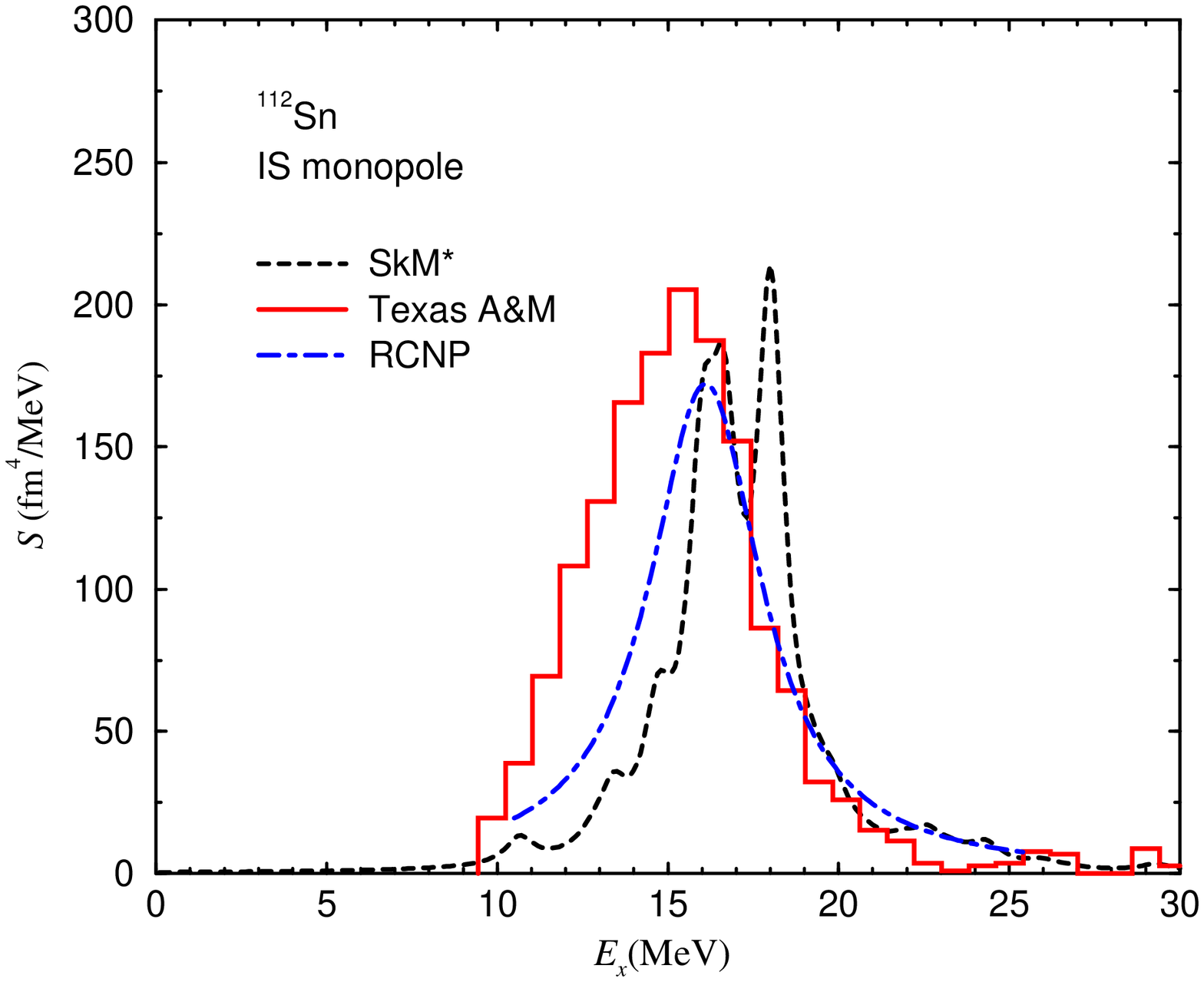}
%\vspace{-4cm}
\includegraphics[width=3.2in,clip]{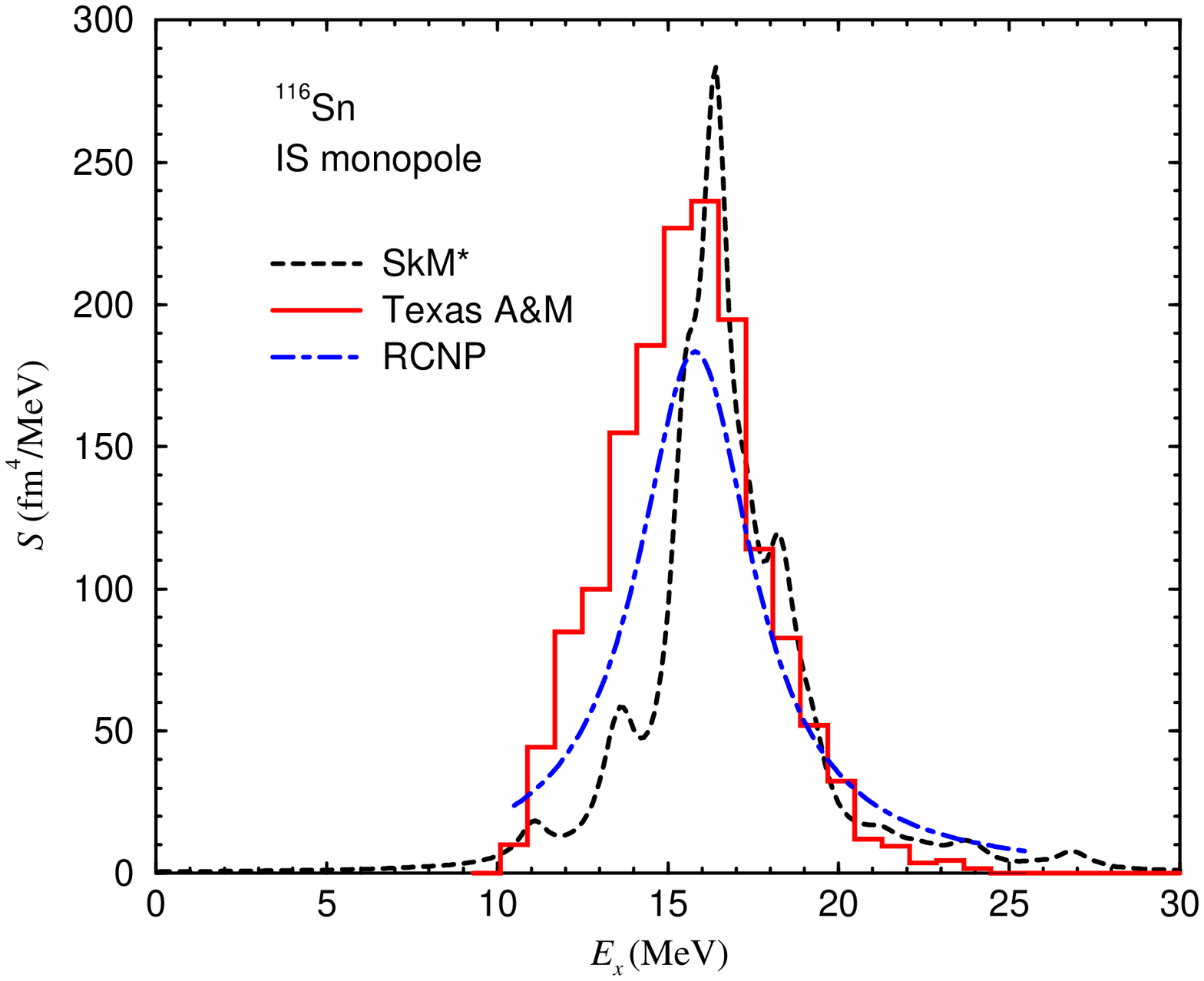}
%\vspace{-3cm}
\includegraphics[width=3.2in,clip]{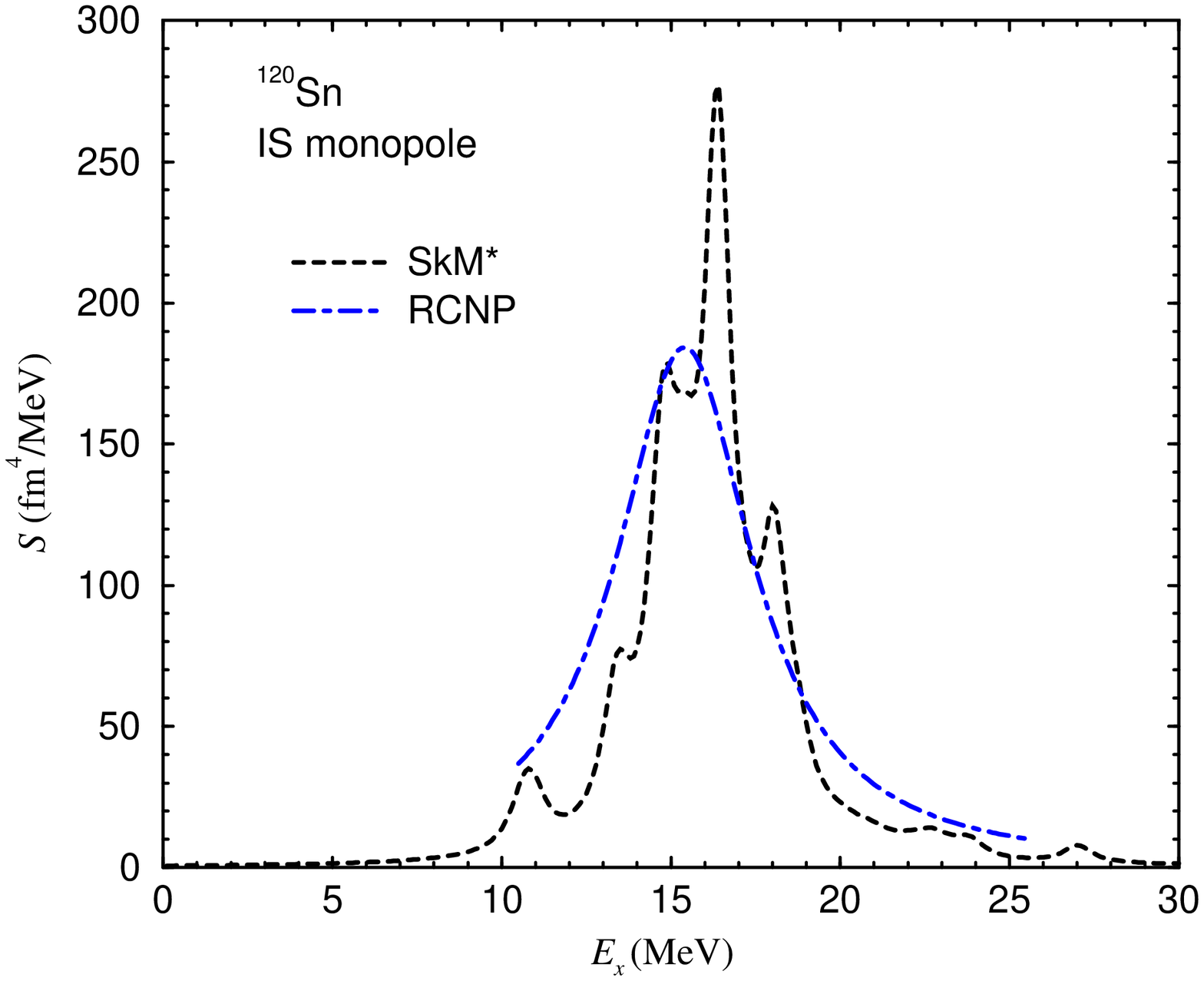}
%\vspace{-3cm}
\includegraphics[width=3.2in,clip]{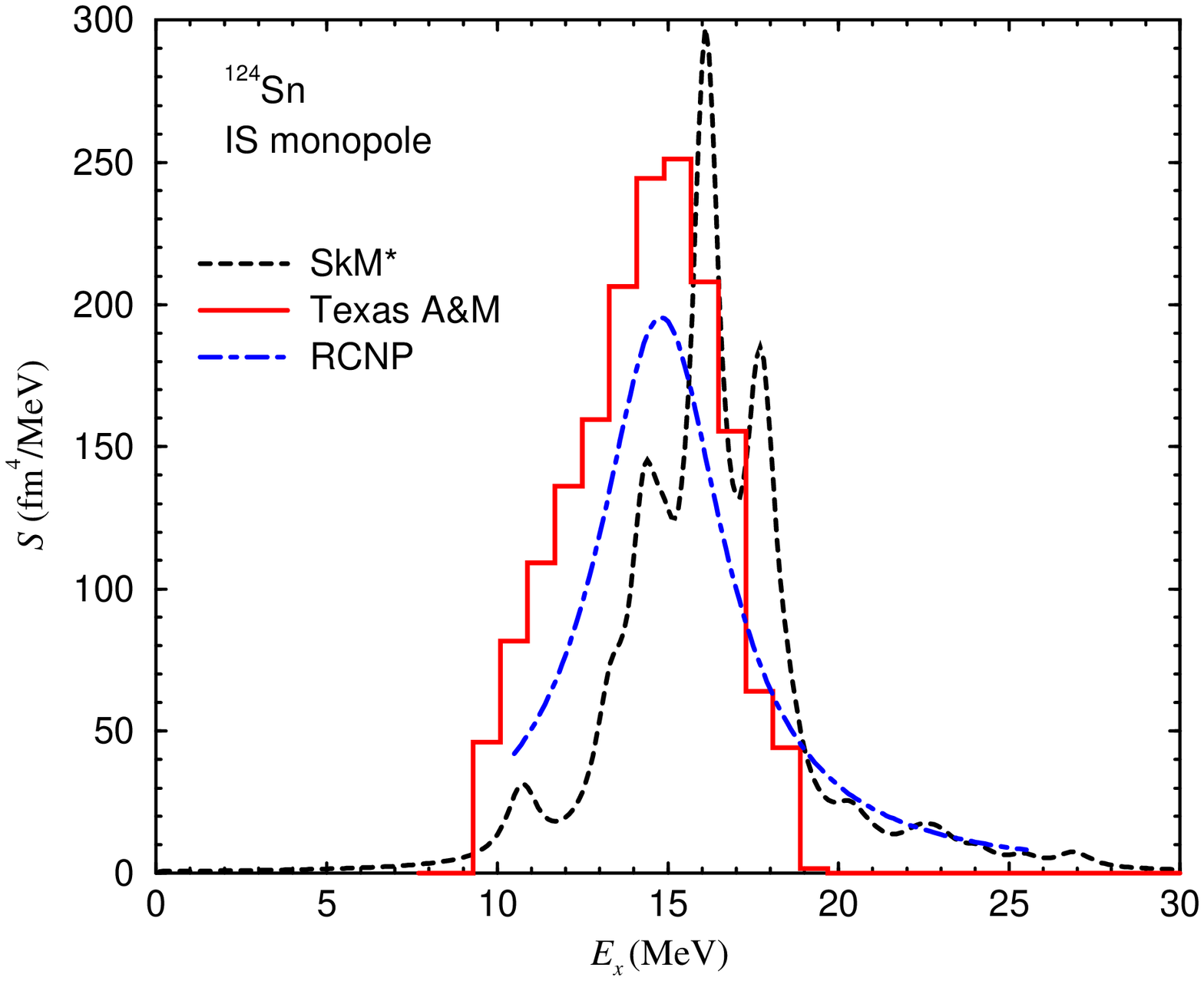}
\caption{(Color online) HF+RPA response functions of $^{112}$Sn, $^{116}$Sn, $^{120}$Sn and 
$^{124}$Sn nuclei with Skyrme interaction
 SkM$^*$.
The experimental data are taken from refs. \cite{Lui} (Texas A\&M) 
 and \cite{Umesh} (RCNP).}
\label{fig:sn-mono}
\end{figure}

\begin{thebibliography}{99}
\bibitem{Skyrme} T. H. R.~Skyrme, Nucl. Phys. {\bf 9}, 615 (1959).
\bibitem{Vautherin} D.~Vautherin and D. M.~Brink, Phys. Rev.~C{\bf 5},
626 (1972).
\bibitem{Blaizot} J.P.~Blaizot, Phys. Rep. {\bf 64}, 171 (1980).
\bibitem{HS97}  I. Hamamoto and H. Sagawa and X.Z.Zhang,  Phys. Rev. C{\bf 56}, 3121(1997).
\bibitem{Serot3} B. D.~Serot, and J. D.~Walecka, Int. Jour. Mod. Phys. E
{\bf 6}, 515 (1997).
\bibitem{Piek02} J. Piekarewicz,  Phys. Rev. C{\bf 66}, 041301(2002).
\bibitem{Vret03} D. Vretenar, T. Nik\u{s}i\'{c}, and P. Ring,  Phys. Rev. C{\bf
68}, 024310(2003).
\bibitem{Yoshida06} S.~Yoshida and H.~Sagawa, Phys. Rev. C{\bf 73}, 044320 
(2006).
\bibitem{yoshida} S.~Yoshida and H.~Sagawa, Phys. Rev. C{\bf 69}, 024318 
(2004).
\bibitem{Brown} B. A.~Brown, Phys. Rev. Lett.~{\bf 85}, 5296 (2000). 
\bibitem{Brown2}
 S. Typel and B. A.~Brown, Phys. Rev. C~{\bf64}, 027302(2001);\\
 R. J. Furnstahl,  Nucl. Phys. {\bf A706},85 (2002).1
\bibitem{Beiner} M.~Beiner, M.~Flocard, N.Van~Giai, and P.~Quentin, Nucl.
Phys.~{\bf A238}, 29 (1975).
\bibitem{Kohler} H. S. ~K\"{o}hler, Nucl. Phys. {\bf A258}, 301 (1976).
\bibitem{Bohigas} O.~Bohigas, H.~Krivine, and J.~Treiner, Nucl. Phys.
{\bf A336}, 155 (1980).
\bibitem{Bartel} J.~Bartel, P.~Quentin, M.~Brack, C.~Guet, and
H.-B.~H\mbox{\aa}kansson, Nucl. Phys. {\bf A386}, 79 (1982).
\bibitem{Reinhard} P.-G.~Reinhard, H.~Flocard, Nucl. Phys.~{\bf A584},
467 (1994).
\bibitem{Chabanat} E.~Chabanat, P.~Bonche, P.~Haensel, J.~Meyer, and
R.~Schaeffer, Nucl. Phys.~{\bf A635}, 231 (1998).
\bibitem{Sharma} M. M. ~Sharma, G.~Lalazissis, J.~K\"{o}nig,
and P.~Ring, Phys. Rev. Lett. {\bf 74}, 3744 (1995).
\bibitem{Giai} Nguyen Van~Giai, H.~Sagawa, Phys. Lett. {\bf 106B}, 379 
(1981).

\bibitem{Toki}Y. Sugawara and H. Toki, Nucl. Phys.~{\bf A579}, 557 (1994).
\bibitem{Lalazissis} G. A.~Lalazissis, J.~K\"{o}nig, and P.~Ring,
Phys. Rev. C{\bf 55}, 540 (1997).
\bibitem{Sharma2} M. M.~Sharma, M. A.~Nagarajan, and P.~Ring, Phys. Lett.
{\bf B312}, 377 (1993).
\bibitem{Nik} T. Nik\v{s}i\'{c}, D. Vretenar, P. Finelli and P.~Ring, 
Phys. Rev. C{\bf 66}, 024306 (2002).
\bibitem{Lala} G. A. Lalazissis, T. Nik\v{s}i\'{c}, D. Vretenar and 
 P.~Ring, Phys. Rev. C{\bf 71},024312 (2005).
\bibitem{Colo} G.~Col\`{o}, N.V.~Giai, J.~Meyer, K.~Bennaceur, and P.~Bonche, 
Phys. Rev. C{\bf 70}, 024307 (2004).
%\bibitem{Danielewicz} P.~Danielewicz, Nucl. Phys. {\bf A727}, 233 (2003).
%\bibitem{Ono}A.~Ono, P.~Danielewicz, W. A.~Friedman, W. G.~Lynch and M. B.~Tsang,
%Phys. Rev. C{\bf 70}, 041604(R) (2004).
%\bibitem{Zamick} L.~Zamick, Phys. Lett. {\bf 45B}, 313 (1973).
%\bibitem{Chabanat2} E.~Chabanat, P.~Bonche, P.~Haensel, J.~Meyer and 
%R.~Schaeffer, Nucl. Phys. {\bf A627}, 710 (1997).
%\bibitem{Cochet} B.~Cochet, K.~Bennaceur, J.~Meyer, P.~Bonche and T.~Duguet, 
%Int. Jour. Mod. Phys. {\bf E13}, 187 (2004).
\bibitem{Youngblood} D. H.~Youngblood, H. L.~Clark, and Y.-W. Lui, Phys. Rev. Lett. 
%{\bf 82}, 691 (1999).
%\bibitem{Uchida} M.~Uchida et al., Phys. Rev. C{\bf 69}, 051301(R) (2004).
%\bibitem{Hoffman} G. W.~Hoffman, {\it etal}, Phys. Rev. {\bf C21}, 1488 (1980).
%\bibitem{Starodubsky} V. E.~Starodubsky and N. M.~Hintz, Phys. Rev. {\bf C49}, 
%2118 (1994).
%\bibitem{Clark} B. C.~Clark, L. J.~Kerr and S.~Hama, Phys. Rev. {\bf C67}, 
%054605 (2003).
%\bibitem{Hof} F.~Hofmann, C. M.~Keil and H.~Lenske, 
%Phys. Rev. C{\bf 64}, 034314 (2001).
%\bibitem{Akmal} A.~Akmal, V. R.~Pandharipande and D. G.~Ravenhall, 
% Phys. Rev. C{\bf 58}, 1804 (1998).
%\bibitem{Friedman} B.~Friedman and V.~R. Pandharipande, Nucl. Phys. {\bf 
%A361}, 502 (1981).
\bibitem{SY}  S. Shlomo and D. H. Youngblood,  Phys. Rev. C{\bf47}, 529 (1993).
\bibitem{Lui}  D. H. Youngblood, Y.-W. Lui, H. L. Clark, B. John, Y. Tokimoto and X. Chen, Lui, Phys. Rev. C{\bf69}, 034315 (2004); Y.-W.Lui, private
  communications. 
\bibitem{Lui2} Y.-W. Lui, D. H. Youngblood, Y. Tokimoto, H.L. Clark, B. John, 
 Phys. Rev. C{\bf70}, 014307(2004).
\bibitem{Umesh} U. Garg, Proc. of 2nd COMEX Meeting (2006) and private 
 communications.\\
 T. Li et al., (preprint,2007).  
 \bibitem{Shlomo}
  T. Sil, S. Shlomo, B. K. Agrawal and P.-G.Reinhard,
  Phys. Rev. C{\bf 73}, 034316(2006).
\end{thebibliography}
\end{document}